\documentclass[12pt]{article}

\usepackage{a4wide,graphics,graphicx,amsmath,amssymb,cite,nicefrac,upgreek}
\usepackage[verbose]{wrapfig}
\usepackage{authblk,hyperref}
\catcode`@=11 \@addtoreset{equation}{section} \catcode`@=12

\begin{document}
\date{}
\title{
{\baselineskip -.2in
%\vbox{\small\hskip 4in \hbox{hep-th/07-10~~~~~~~~~~}}
%\vbox{\small\hskip 4in \hbox{IITM/PH/TH/2014/2}}
} 
\vskip .4in
\vbox{
{\bf \LARGE Self-supporting wormholes with massive vector field}
}}

\author{Ankit Anand and Prasanta K. Tripathy\thanks{email: Anand@physics.iitm.ac.in, prasanta@iitm.ac.in}}
\affil{\normalsize\it Department of Physics, \authorcr \it Indian Institute of Technology Madras, \authorcr \it Chennai 600036, India.}
\maketitle
\begin{abstract}

In this paper we consider a massive vector field in the background of a space-time obtained by certain $\mathbb{Z}_2$ quotient of the 
BTZ black hole. We analyse the back reaction of the matter field on the space-time geometry up to first order in metric perturbation. The 
expectation value of the stress-energy tensor can be computed exactly by considering its pull-back onto the covering space. 
Upon a suitable choice of the boundary condition on the vector field around a non-contractible cycle of the quotient manifold it is 
possible to obtain the average energy on a null geodesic to be negative there by resulting a traversable wormhole.

\end{abstract}

\newpage
\section{Introduction}

Wormholes are solutions to Einstein equations which connect two otherwise distinct space-times or two widely separated regions of the 
same space-time via a throat. In classical general relativity wormholes are not traversable, {\it i.e.}, no causal curve can pass through the 
throat of the wormhole connecting the two distinct regions. The issue of traversability for static, spherically symmetric wormholes has been 
discussed first in \cite{Morris:1988cz} where it has been pointed out the need to have exotic matters for the wormhole to be traversable.
This matter has further been analysed in \cite{Hochberg:1998ha,Morris:1988tu,Visser:2003yf} substantiating the violation of average null 
energy condition (ANEC) as the necessary condition for traversability of wormhole. It has been proved that the ANEC holds for achronal 
null geodesics \cite{Graham:2007va,Wall:2009wi,Kelly:2014mra}. Thus, space-times which possess only achronal null geodesics do not 
admit traversable wormholes. These results have their origin lies in the topological censorship theorem \cite{Friedman:1993ty} and its 
generalisation to asymptotically locally anti-de Sitter spaces \cite{Galloway:1999bp}, which states that every causal curve whose end 
points lie in the boundary at infinity (${\cal I}$) can be deformed to a causal curve which entirely lies in $\cal I$ itself.

An important breakthrough in this direction has recently been achieved by Gao, Jafferis and Wall \cite{Gao:2016bin} where they have 
constructed a traversable wormhole from an eternal BTZ black hole by introducing a time-dependent coupling between its two asymptotic 
regions. They have computed the one loop stress energy tensor upon using the point splitting method. By choosing the sign of the coupling 
appropriately the vacuum expectation value of the double null component of the stress energy tensor can be made negative enabling 
the wormhole traversable. These results have subsequently been generalised in \cite{Caceres:2018ehr} to see the effect of rotation on
the size of the wormhole. An eternal traversable wormhole in nearly-$AdS_2$ space-time has been constructed in \cite{Maldacena:2018lmt} by introducing a coupling between the two boundaries. 

A traversable wormhole in four dimensions has been devised in \cite{Maldacena:2018gjk} by joining the throats of two charged 
extremal black hole geometries in the presence of massless fermions. This construction did not depend on any non-local external coupling between
the two boundaries and resulted to what are called as the self-supporting wormholes  which arise entirely from local dynamics of the 
fermion fields present in the bulk of the space-time. A complementary analysis has been carried out in \cite{Fu:2018oaq} to obtain traversable wormholes from the bulk dynamics 
by considering a free scalar field in quotients of $AdS_3$ and $AdS_3\times S^1$ by discreet symmetries. The authors computed the 
gravitational back reaction and showed that the space-time admits causal curves which can't be deformed to the boundary. Taking quotient  by a discrete symmetry is significant in the sense that,
it no longer preserves the globally defined Killing field which plays a key role in obtaining the 
average null energy condition. This result 
has subsequently been generalised \cite{Marolf:2019ojx} to include fermions in the bulk to produce traversable wormholes.

In the present work we generalise these above results in the presence of massive vector fields. As 
has been noticed in \cite{Marolf:2019ojx}, adding spin-half fields provide a rich structure which is 
worth studying in its own right. We will notice similar phenomenon in the presence of spin-one fields.
In what follows, we consider the pullback of the stress tensor on to the covering space. For $AdS_3$ it is possible
to obtain an exact analytic expression for the propagator in closed form. Using this result we compute 
the expectation value of the stress tensor and show that under imposing suitable boundary 
conditions, this gives rise to traversable wormholes. The plan of the paper is as follows. The next
section summarises the preliminaries on obtaining self-supporting wormholes from free scalar 
fields. In \S3 we obtain the exact propagator in $AdS_3$. Subsequently we compute the expectation 
value of the stress tensor from this propagator by using the method of images. The appendix discusses 
some of the technical details.

\section{Preliminaries}

Consider the $AdS_3$ metric in Kruskal-like coordinates:
\begin{equation}
dS^2 = \frac{1}{\left(1 + U V\right)^2} \Big( - 4\ell^2 dU dV + r_+^2 \left(1 - U V\right)^2 d\varphi^2\Big) \ .
\end{equation}
This gives rise to non-rotating BTZ black hole with horizon radius $r_+$ upon imposing the identification $\varphi\sim\varphi+2\pi$ for the
azimuthal angle.  The horizons of the black hole are located at $U=0$ and $V=0$ respectively. The space-time boundaries correspond to 
$1 + U V =0$. This solution has been discussed from the perspective of gauge-gravity duality in \cite{Maldacena:2001kr}. The issue of 
traversability in this geometry has been analysed first in \cite{Gao:2016bin} by considering a relevant double trace deformation coupling the 
two boundaries. This relevant deformation in the boundary CFT amounts to adding a stress tensor in the bulk resulting a perturbation of the 
space-time geometry. 

Consider the $V=0$ horizon which admits the horizon generator $k^\rho$ such that $k^\rho\partial_\rho = \partial_U$. We choose $U$ to 
be the affine parameter which parametrises the null geodesics tangent to this horizon. The linearised Einstein's equation for the metric 
perturbation on the $V=0$ horizon is given by:
\begin{equation}\label{linee} 
\frac{1}{2\ell^2}\left(h_{UU} + \partial_U(U h_{UU}) - \frac{\ell^2}{r_+^2} \partial_U^2 h_{\varphi\varphi}\right)
 = 8\pi G_N  T_{UU} \ .
 \end{equation}
The geodesic equation for null rays originating on the past horizon on the other hand gives rise to 
\begin{equation}
\Delta V(U) = \frac{1}{2\ell^2} \int_{-\infty}^U dU\ h_{UU} \ .
\end{equation}
The quantity $\Delta V(\infty)$ measures the time delay of the null geodesic starting at $U=-\infty$ and ending at $U=\infty$. The wormhole 
becomes traversable if $\Delta V(\infty)<0$. This quantity provides a measure for the size of the opening of the wormhole. 
Integrating \eqref{linee} over $U$ keeping in mind the perturbation vanish at the boundary gives 
rise to 
\begin{equation}
\Delta V(\infty) = 8\pi G_N \int_{-\infty}^\infty dU\ T_{UU} \ .
\end{equation}
Thus the wormhole becomes traversable if the ANEC is violated. By choosing suitable non-local coupling between the two boundaries it 
has been shown \cite{Gao:2016bin} that it is indeed possible to violate the ANEC giving rise to traversable wormholes.

An alternative method has been developed in \cite{Fu:2018oaq} to construction traversable wormholes without invoking any non-local
boundary interaction. The authors considered a suitable $\mathbb Z_2$ quotient of the $BTZ$ black hole space-time $\tilde M$. The 
resulting geometry $M$ is a smooth, globally hyperbolic manifold, known as the $\mathbb{RP}^2$-geon \cite{Louko:1998hc}. 
The $\mathbb Z_2$ quotient introduces a new homotopy cycle in the manifold $M$. One 
can choose the scalar field in $M$ to be either periodic or anti-periodic around this cycle.  The 
states in $M$ are constructed from the states of the covering space $\tilde M$ by the method of images. This 
enables the average expectation value in a suitable Hartle-Hawking like state of the double null 
component of the resulting stress tensor to take negative values along null geodesics.

Let $J$ be the isometry which maps the point $\tilde x\in \tilde M$ to $J\tilde x$, and the pair 
$(\tilde x, J\tilde x)$ projects onto the point $x$ in the quotient $M$. The quantum fields 
$\phi_\pm(x)$ in $M$ are constructed from the quantum fields $\tilde\phi(\tilde x)$ in $\tilde M$ 
using the method of images:

\begin{equation}
\phi_\pm(x) = \frac{1}{\sqrt 2}\Big( \tilde\phi(\tilde x) \pm \tilde\phi(J\tilde x)\Big) \ .
\end{equation}
The pair of points $(\tilde x, J\tilde x)$ are space-like separated in $\tilde M$ and hence the fields 
at these two points commute with each other. Thus, the fields $\phi_\pm(x)$ satisfies the 
usual canonical commutation relations and describe well defined quantum fields in $M$.

We consider free scalar field $\phi_\pm(x)$ in $M$ for which the action is given by
\begin{equation}
S = \int d^3x \sqrt{-g} \left( -\frac{1}{2} g^{\mu\nu} \partial_\mu\phi_\pm \partial_\nu\phi_\pm - \frac{1}{2} m^2 \phi_\pm^2 \right) \ ,
\end{equation}
with the corresponding energy-momentum tensor 
\begin{equation}
T_{\mu\nu\pm} = \partial_\mu\phi_\pm \partial_\nu\phi_\pm - \frac{1}{2} g_{\mu\nu} \left(
g^{\alpha\beta}\partial_\alpha\phi_\pm\partial_\beta\phi_\pm + m^2\phi_\pm^2\right) \ . 
\end{equation}
The Hartle-Hawking state $|HH,\tilde M\rangle$ in the covering space $\tilde M$ induces corresponding states $|HH,\pm\rangle$ in $M$. 
We need to compute the expectation value of the double null component of the stress tensor 
$\langle HH,\pm | T_{\mu\nu}k^\mu k^\nu(x)|HH,\pm \rangle$ in the Hartle-Hawking states $|HH,\pm\rangle$. It is most convenient
to compute this by considering the pullback $T^P_{\mu\nu}(\tilde x)$ in $\tilde M$ of the stress-tensor $T_{\mu\nu}(x)$:
\begin{equation}
\langle HH,\pm | T_{\mu\nu\pm}k^\mu k^\nu(x)|HH,\pm \rangle 
= \langle HH,\tilde M | T^P_{\mu\nu\pm}k^\mu k^\nu(\tilde x)|HH,\tilde M \rangle \ .
\end{equation}
It is important to note that $T^P_{\mu\nu}(\tilde x)$ is not the stress-tensor of a free quantum field $\tilde\phi(\tilde x)$ in $\tilde M$. 
Since the Hartle-Hawking state  $|HH,\tilde M\rangle$ is invariant under the Killing symmetry, the expectation value of the stress tensor
$\tilde T_{\mu\nu}(\tilde x)$ in this state vanishes. This is because, the expectation value of $\xi^\mu \xi^\nu \tilde T_{\mu\nu}(\tilde x)$ is 
invariant for a Killing vector $\xi^\mu$. If $\xi^\mu$ becomes $k^\mu$ on the horizon then this quantity vanishes on the bifurcation surface, 
and hence the expectation value of the double null component of the stress tensor for the covering space must vanish identically. However,
this need not be the case for the pullback $T^P_{\mu\nu}(\tilde x)$, as this quantity is not preserved by the Killing symmetry. A straightforward 
analysis gives rise to 
\begin{equation}
T^P_{\mu\nu\pm}k^\mu k^\nu(\tilde x) 
= \tilde T_{\mu\nu}k^\mu k^\nu(\tilde x)  \pm k^\mu\partial_\mu\tilde\phi(\tilde x) k^\nu\partial_\nu\tilde\phi(J\tilde x) \ .
\end{equation}
Thus, we have 
\begin{equation}
\langle HH,\pm | T_{\mu\nu\pm}k^\mu k^\nu(x)|HH,\pm \rangle  =
 \pm \langle HH,\tilde M| k^\mu\partial_\mu\tilde\phi(\tilde x) k^\nu\partial_\nu\tilde\phi(J\tilde x) |HH,\tilde M\rangle \ .
\end{equation}
Hence, if the expectation value on the right hand side is non-zero, we can always choose appropriate boundary condition to make it 
negative. This has been computed for various smooth, globally hyperbolic, $\mathbb Z_2$ quotients of $BTZ$ and $BTZ\times S^1$
space-times \cite{Fu:2018oaq}. The results have been used thereafter to show the violation of ANEC.

\section{The massive vector fields}

We will now consider the case of an Abelian vector field of mass $m$ with the action
\begin{equation}\label{actionone}
S = \int d^3x \sqrt{-g} \left( - \frac{1}{4} g^{\alpha\beta} g^{\mu\nu} F_{\alpha\mu} F_{\beta\nu} 
- \frac{1}{2} m^2 g^{\mu\nu} A_\mu A_\nu \right)
\end{equation}
As in the case of scalar fields, we can use the method of images to set the vector field $A_{\mu}(x)$ in $M$, in terms of the corresponding
fields $\tilde A_\mu(\tilde x)$ in the covering space $\tilde M$ as:
\begin{equation}
A_{\mu}^{\pm}(x) = \frac{1}{\sqrt 2} \Big( \tilde A_\mu(\tilde x) \pm \tilde A_{\mu'}(\tilde x') \Big) \ . 
\end{equation}  
The $(\pm)-$sign in the superscript correspond to the choice of periodic or anti-periodic boundary condition on the vector field 
$A_\mu(x)$ around the non-trivial homology cycle $\gamma$. The above setting will give rise to a well defined quantum field in $M$ 
for space-like separated points $(\tilde x, \tilde x')$. Here we will set $\tilde x' = J\tilde x$. The pair of points $(\tilde x, J\tilde x)$ in 
$\tilde M$ maps onto the point $x \in M$ under the action of $\mathbb Z_2$. 

The stress tensor corresponding to the action \eqref{actionone} is given by
\begin{equation}
T_{\mu\nu}^{\pm}(x) = g^{\alpha\beta} F_{\alpha\mu}^{\pm} F_{\beta\nu}^{\pm} + m^2 A_{\mu}^{\pm} A_{\nu}^{\pm} 
- g_{\mu\nu} \left( \frac{1}{4} g^{\alpha\beta} g^{\rho\sigma} F_{\alpha\rho}^{\pm} F_{\beta\sigma}^{\pm} 
+ \frac{1}{2} m^2 g^{\alpha\beta} A_{\alpha}^{\pm} A_{\beta}^{\pm}\right) \ .
\end{equation}
The last term in the above equation does not contribute to the double null component of $T_{\mu\nu}$. Hence, we will ignore this term
now on. The pullback of the first two terms into the covering space $\tilde M$ gives rise to 
\begin{eqnarray}
k^\mu k^\nu T^{\pm}_{\mu\nu}(x) = \frac{1}{2}\Big(k^\mu k^\nu  \tilde T_{\mu\nu}(\tilde x) +  k^{\mu'} k^{\nu'}  \tilde T_{\mu'\nu'}(\tilde x') \Big) 
\pm \frac{1}{2} \left\{  \Big( g^{\alpha\beta'}(\tilde x,\tilde x')k^\mu \tilde F_{\alpha\mu}(\tilde x)  k^{\nu'} \tilde F_{\beta'\nu'}(\tilde x') \right. \cr 
+   \left. g^{\alpha'\beta}(\tilde x',x)k^{\mu'} \tilde F_{\alpha'\mu'}(\tilde x')  k^\nu \tilde F_{\beta\nu}(\tilde x) \Big)
+ m^2 \Big( k^\mu \tilde A_\mu(\tilde x) k^{\nu'} \tilde A_{\nu'}(\tilde x') +  k^{\mu'} \tilde A_{\mu'}(\tilde x') k^\nu\tilde A_\nu(\tilde x)\Big) 
\right\}  .
\end{eqnarray}
Here $\tilde T_{\mu\nu}(\tilde x)$ is the stress tensor corresponding to the gauge field $\tilde A_\mu(\tilde x)$ in $\tilde M$. The 
expectation values of $T_{\mu\nu}(x)$ in the states $|HH,\pm\rangle$ become
\begin{eqnarray} \label{pullback}
\langle HH,\pm| k^\mu k^\nu T^{\pm}_{\mu\nu}(x)  |HH,\pm\rangle &=&\pm \left\{  \langle HH,\tilde M| \Big(
g^{\alpha\beta'}(\tilde x,\tilde x')k^\mu \tilde F_{\alpha\mu}(\tilde x) k^{\nu'} \tilde F_{\beta'\nu'}(\tilde x') 
\Big) | HH,\tilde M\rangle \right. \cr 
&+& \left. m^2  \langle HH,\tilde M| \Big( k^\mu \tilde A_\mu(\tilde x) k^{\nu'} \tilde A_{\nu'}(\tilde x') \Big) |HH,\tilde M\rangle \right\} \ .
\end{eqnarray}
The expectation value $\langle HH,\tilde M| k^\mu k^\nu  \tilde T_{\mu\nu}(\tilde x)|HH,\tilde M\rangle$ vanishes because of symmetry
as argued in the previous section. Again, as in the scalar field case, we can impose appropriate boundary conditions on the vector field to 
make the null energy negative provided the right hand side in \eqref{pullback} does not vanish.

\subsection{The propagator}

From the above analysis we find that, in order to verify the ANEC, we need to compute the 
expectation values $ \langle HH,\tilde M|  \tilde A_\mu(\tilde x) \tilde A_{\nu'}(\tilde x') |HH,\tilde M\rangle $ and 
$\langle HH,\tilde M| \tilde F_{\alpha\mu}(\tilde x)F_{\beta'\nu'}(\tilde x') | HH,\tilde M\rangle$ in
the covering space $\tilde M$. The manifold $\tilde M$ itself can be obtained upon taking the
quotient of $AdS_3$ with the identification $\varphi \sim \varphi + 2\pi$. Thus, we can derive 
these quantities from the vector two-point function in $AdS_3$.

The vector two-point function in $AdS_n$ in arbitrary dimensions has been computed in \cite{Allen:1985wd}. In the following we will briefly 
outline the relevant parts of their results for our purpose. Denote $A_\mu(x)$ to be the massive vector field in $AdS_n$. The two-point 
function $Q_{\mu\nu'}(x,x') = \langle A_\mu(x) A_{\nu'}(x')\rangle$ evaluated on the vacuum is a maximally symmetric bitensor in $AdS_n$. 
On general grounds it can be express  as 
\begin{equation} \label{qalfbet}
Q_{\alpha\beta'}(x,x') =  \upalpha(\upmu) g_{\alpha\beta'}(x,x') + \upbeta(\upmu) n_\alpha(x,x') n_{\beta'}(x,x') \ .
\end{equation}
Here $\upmu(x,x')$ is the geodesic distance between the points $x$ and $x'$, the unit vectors $n_\alpha$ and $n_{\alpha'}$ are the 
covariant derivatives of $\upmu$ with respect to $x^\alpha$ and $x^{\alpha'}$ respectively:
\begin{equation} \label{ptrpt}
n_\alpha(x,x') = \nabla_\alpha\upmu(x,x') \ , \ n_{\alpha'} = \nabla_{\alpha'} \upmu(x,x') \ , 
\end{equation}
and $g_{\alpha\beta'}(x,x')$ is the parallel propagator along the geodesic joining $x$ and $x'$. This quantity is uniquely defined as the linear map
which parallel transports vectors along the geodesics. The equation of motion for the parallel propagator can be obtained from the equation 
for parallel transport of a vector and is given by:
\begin{equation}\label{ppgeom}
\frac{d}{d\lambda} {g^\mu}_{\nu'}(x,x') + \Gamma^\mu_{\nu\rho} \frac{dx^\nu}{d\lambda}{g^\rho}_{\nu'}(x,x') = 0 \ .
\end{equation}

The functions $\upalpha(\upmu)$ and $\upbeta(\upmu)$ in \eqref{qalfbet} are determined by requiring that 
$Q_{\mu\nu'}(x,x')$ satisfies the equation of motion and by examining its singularity structure. They, in turn are determined in terms of a 
function $\upgamma(\upmu)$ as 
\begin{eqnarray} \label{alphab}
\upalpha(\upmu) = \upbeta(\upmu) +\upgamma(\upmu) = \frac{\ell}{(n-1)}  \sinh\left(\nicefrac{\upmu}{\ell}\right) \upgamma'(\upmu) + \cosh\left(\nicefrac{\upmu}{\ell}\right)\upgamma(\upmu) \ . 
\end{eqnarray}
The function $\upgamma(\upmu)$ is expressed in terms of hypergeometric functions 
\begin{equation}
\upgamma(z) = r\ z^{-a_+} F\left(a_+,a_+ - c +1;a_+ - a_- +1;z^{-1}\right) \ ,
\end{equation}
with $z = \cosh^2\left(\nicefrac{\upmu}{2\ell}\right)$. The parameters $a_\pm$ and $c$ are given by
\begin{eqnarray}
2 a_\pm =  (n+1) \pm \sqrt{(n-3)^2 + 4 m^2\ell^2} \ , \ {\rm and} \ 
2 c = n + 2 \ .
\end{eqnarray}
The normalisation factor $r$ is given by
\begin{equation}
r = \frac{(1-n) \Gamma(a_+)\Gamma(a_+-c+1)}{2^{n+1}\pi^{n/2} m^2\ell^n \Gamma(a_+-a_-+1)} \ .
\end{equation}

For $AdS_3$ the expression for $\upgamma(z)$ takes the form
\begin{equation}
\upgamma(z) = - \frac{2 (m\ell + 1)}{m^2\ell^3 \pi (4 z)^{m\ell + 2}} F\left(m\ell + 2, m\ell + 1/2; 2 m\ell + 1; z^{-1}\right) \ .
\end{equation}
This hypergeometric series can be summed to obtain an exact analytic expression in closed form for the function $\upgamma(z)$. 
We find 
\begin{equation}
\upgamma(z) = - \frac{1}{16\pi m^2\ell^3}
 \frac{\left(\ 2z - 1 + 2 m\ell \sqrt{ z (z-1)} \ \right) }
 {z^{3/2} \big(z-1\big)^{3/2} \big(\sqrt{z} + \sqrt{z-1}\big)^{2m\ell}} \ .
\end{equation}
As expected, this quantity has the usual singularities at $z=0,1$ and branch cut along the real axis for $z < 1$. Substituting 
$z = \cosh^2\left(\nicefrac{\upmu}{2\ell}\right)$, we find $\upgamma(z)$ as a function of $\upmu$ has the simple expression:
\begin{equation}
\upgamma(\upmu) = - \frac{1}{2\pi m^2\ell^3} \frac{e^{-m \upmu(x,x')}}{\sinh^3\left(\nicefrac{\upmu}{\ell}\right) }
\Big(\cosh\left(\nicefrac{\upmu}{\ell}\right) + m\ell \sinh\left(\nicefrac{\upmu}{\ell}\right)\Big) \ .
\end{equation}
Substituting the above for $\upgamma(\upmu)$ in \eqref{alphab}, we find that the two-point function for the massive vector 
field $A_\mu(x)$ in $AdS_3$ has the form
\begin{equation} \label{twopf}
\langle A_\delta(x) A_{\sigma'}(x')\rangle =  \upalpha(\upmu) g_{\delta\sigma'}(x,x') + \upbeta(\upmu) n_\delta(x,x') n_{\sigma'}(x,x') \ 
\end{equation}
where the functions $\upalpha(\upmu)$ and $\upbeta(\upmu)$ are given by
\begin{eqnarray}\label{abgfmla}
\upalpha(\upmu) &=& \frac{e^{-m \upmu(x,x')} }{4\pi m^2\ell^3} \rm{cosech}\left(\nicefrac{\upmu}{\ell}\right)
\Big( m^2\ell^2 + m\ell \coth\left(\nicefrac{\upmu}{\ell}\right) +  \rm{cosech}^2\left(\nicefrac{\upmu}{\ell}\right) \Big) \\
\upbeta(\upmu) &=&  \frac{e^{-m \upmu(x,x')} }{4\pi m^2\ell^3} \rm{cosech}\left(\nicefrac{\upmu}{\ell}\right)
\Big( \coth\left(\nicefrac{\upmu}{\ell}\right) \big(m\ell + 2 \rm{cosech}\left(\nicefrac{\upmu}{\ell}\right)\big) \nonumber
+ \big(m\ell + \rm{cosech}\left(\nicefrac{\upmu}{\ell}\right)\big)^2 \Big)
\end{eqnarray}

Using the techniques involving bitensors developed in \cite{Allen:1985wd} the two-point function involving the field strengths can be 
calculated from the above in a straightforward manner. Some of the intermediate steps are outlined in appendix ${\bf A}$. We find 
\begin{eqnarray} \label{twofc}
\langle F_{\eta\delta}(x) F_{\rho'\sigma'}(x')\rangle 
&=& m^2 \upbeta(\upmu) \big(g_{\delta\sigma'} n_\eta n_{\rho'} - g_{\eta\sigma'} n_\delta n_{\rho'}
+ g_{\eta\rho'} n_\delta n_{\sigma'}  - g_{\delta\rho'} n_\eta n_{\sigma'} \big) \cr 
&+& m^2 \upgamma(\upmu) \big( g_{\eta\sigma'} g_{\delta\rho'} - g_{\delta\sigma'} g_{\eta\rho'} \big) 
\end{eqnarray}
Here as an aside we note that:
\begin{eqnarray}
\lim_{m\rightarrow 0} m^2 \upbeta(\upmu) = \frac{1 + 2 \cosh\left(\nicefrac{\upmu}{\ell}\right)}{4\pi\ell^3 \sinh^3\left(\nicefrac{\upmu}{\ell}\right)}
\ , \ {\rm and} \ \lim_{m\rightarrow 0} m^2 \upgamma(\upmu) = - \frac{\cosh\left(\nicefrac{\upmu}{\ell}\right)}{2\pi\ell^3 \sinh^3\left(\nicefrac{\upmu}{\ell}\right)} \ .
\end{eqnarray}
Thus, although the right hand side in \eqref{twopf} diverges in the limit $m\rightarrow 0$, the two point 
function involving the field strengths as given above is well defined. 
We will now turn our attention to the quantity of interest for our purpose. From \eqref{twofc} we find 
\begin{equation}
\langle g^{\eta\rho'} F_{\eta\delta}(x) F_{\rho'\sigma'}(x')\rangle = m^2\upbeta(\upmu) \big(n_\delta n_{\sigma'} - g_{\delta\sigma'}\big)
- 2 m^2\upgamma(\upmu) g_{\delta\sigma'} \ . 
\end{equation}
Combining the above with \eqref{twopf} we find
\begin{equation}\label{adsnull}
\langle \big(g^{\eta\rho'} F_{\eta\delta}(x) F_{\rho'\sigma'}(x') + m^2 A_\delta(x) A_{\sigma'}(x')\big)\rangle = 
2 m^2\upbeta(\upmu) n_\delta n_{\sigma'} - m^2\upgamma(\upmu) g_{\delta\sigma'} \ .
\end{equation}

\subsection{The average null energy}

We will now compute the average null energy. We consider the non-rotating BTZ black hole geometry. The metric is given by
\begin{equation}
dS^2 = \frac{1}{\left(1 + U V\right)^2} \Big( - 4\ell^2 dU dV + r_+^2 \left(1 - U V\right)^2 d\varphi^2\Big) \ .
\end{equation}
The co-ordinate $\varphi$ is periodic with the identification $\varphi \sim \varphi + 2\pi$. The $\mathbb{RP}^2$-geon \cite{Louko:1998hc} is 
obtained upon taking the quotient of this geometry with the $\mathbb Z_2$ isometry $J$ which acts on the co-ordinates as 
$J: \left(U, V, \varphi\right) \longrightarrow \left(V, U, \varphi + \pi\right)$. The resulting space-time is a time-orientable manifold with constant 
time hypersurfaces of topology $\mathbb{RP}^2\backslash\{ \rm{point\ at\ infinity}\}$. It gives rise to a black hole with a single exterior and is 
locally identical to the BTZ geometry. 

The two-point function in the Hartle-Hawking state of the $BTZ$ black hole  is obtained from the corresponding two-point function in
$AdS_3$ vacuum by using the method of images with periodic boundary condition. Thus, to get the two-point function we replace 
$\varphi'$ by $\varphi' + 2\pi n$ and sum over all integer values of $n$:
\begin{equation}
%\langle HH,BTZ|A_\rho(x) A_{\sigma'}(x')|HH,BTZ\rangle 
\langle A_\rho(x) A_{\sigma'}(x') \rangle
= \sum_{n\in \mathbb Z} \left( \upalpha(\upmu(x,x'_n)) g_{\rho\sigma'}(x,x'_n) + \upbeta(\upmu(x,x_n')) n_\rho(x,x'_n) n_{\sigma'}(x,x'_n) \right) \ ,
\end{equation}
where $x_n' = (U',V',\varphi_n')$ with $\varphi_n' = \varphi' + 2\pi n$. From now on, we evaluate the expectation values in the 
Hartle-Hawking state. For the $\mathbb{RP}^2$-geon $\varphi' = \varphi + \pi$ and hence $\varphi'_n = \varphi + (2 n +1) \pi$.
Likewise, we have [with $F(x,x')\equiv \left(2 \upbeta(\upmu) n_\delta n_{\sigma'} - \upgamma(\upmu) g_{\delta\sigma'} \right)$]:
\begin{equation}\label{ads-null}
\langle \big(g^{\eta\rho'} F_{\eta\delta}(x) F_{\rho'\sigma'}(x') + m^2 A_\delta(x) A_{\sigma'}(x')\big)\rangle = 
\sum_{n\in\mathbb Z} m^2 F(x,x_n')  \ .
\end{equation}

To compute the above, note that the geodesic distance $\upmu(x,x')$ between points $x$ and $x'$ in $AdS_3$ is given by
\begin{equation}
\cosh\left(\nicefrac{\upmu}{\ell}\right)  = \frac{
\Big( 2 \left( U V' + V U'\right) + \left(1 - U V\right) \left(1 - U' V'\right) \cosh\left(\nicefrac{r_+(\varphi - \varphi')}{\ell}\right)\Big)
}{\left(1 + U V\right) \left(1 + U' V'\right)} \ . 
\end{equation}
It is straightforward to compute the unit vectors $n_\mu(x,x')$ and $n_{\mu'}(x,x')$ from the above expression. This has been carried out in 
appendix $\bf B$. On the $V=0$ surface, they have the form
\begin{eqnarray} \label{univecs} 
n_\mu &=&  \frac{2\ell}{\sinh\left(\nicefrac{\mu}{\ell}\right)} 
\Big( U, - U^3 - U \cosh\left(\nicefrac{k \pi r_+}{\ell}\right), - \frac{r_+}{2\ell} \sinh\left(\nicefrac{k \pi r_+}{\ell}\right) \Big) \ , \cr
n_{\mu'} &=&  \frac{2\ell}{\sinh\left(\nicefrac{\mu}{\ell}\right)} 
\Big( - U^3 - U \cosh\left(\nicefrac{k \pi r_+}{\ell}\right), U, \frac{r_+}{2\ell} \sinh\left(\nicefrac{k \pi r_+}{\ell}\right) \Big) \ .
\end{eqnarray}

We now need to compute the parallel propagator $g(x,Jx)$ on the $V=0$ surface. We can choose $U$ to be the affine parameter along
the geodesic. Now, analysing the equation of motion for the parallel propagator \eqref{ppgeom} it can be shown that the components 
$g_{\mu\nu'}(x,x')$ are independent of $U$ on the $V=0$ surface (see appendix ${\bf C}$ for the derivation).
Since the tangent vectors $n_\alpha(x,x')$ and $n_{\alpha'}(x,x')$ are oppositely directed, we must have:
\begin{equation} \label{ppgrln}
g_{\alpha\beta'}(x,x') n^{\beta'}(x,x') + n_\alpha(x,x') = 0 \ .
\end{equation}
 Requiring $g_{\mu\nu'}$ 
to satisfy the relation this can be used to find 
\begin{eqnarray} \label{ppgfnl}
g(x,x')\biggr\rvert_{\substack{ x'=Jx \\ V=0 }} = \left(\begin{matrix} 2\ell^2 &0 & 0 \cr 0 & 2\ell^2 & 0 \cr 0 & 0  & r_+^2 \end{matrix}\right) \ .
\end{eqnarray}

\begin{figure}
\fbox{\includegraphics[height=3in]{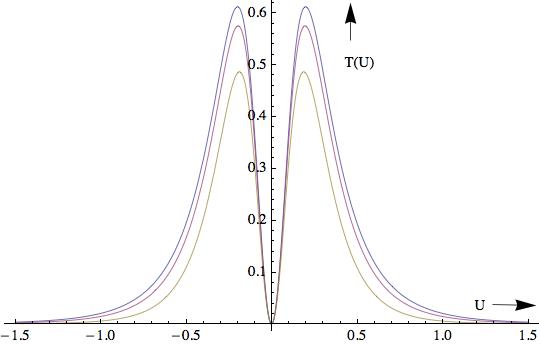}}
\caption{ Variation of $T(U)$ as a function $U$ for $m=0,1$ and $2$ respectively. We have set $\ell=1$ and choose $r_+=1/2\pi$.
The maximum value of $T(U)$ decreases with increasing $m$.}
 \label{tofua}
\end{figure}

For easy reading, we introduce the notation $T(U)$ to denote the null energy:
\begin{equation} \label{nener}
T(U) = \langle k^\delta k^{\sigma'} \big(g^{\eta\rho'} F_{\eta\delta}(x) F_{\rho'\sigma'}(x') + m^2 A_\delta(x) A_{\sigma'}(x')\big)\rangle\biggr\rvert_{\substack{ x'=Jx \\ V=0 }}  \ .
\end{equation}
Note that, $k^\rho$ is the horizon generator on $V=0$. Thus, we must have $k^\rho = (1,0,0)$. Now, substituting  \eqref{adsnull} , \eqref{univecs} and \eqref{ppgfnl} in the above equation we find 
\begin{equation}
T(U) = 8 m^2\ell^2 \frac{\upbeta(\upmu_0) U^2}{\sinh^2\left(\nicefrac{\upmu_0}{\ell}\right)} \ , 
\end{equation}
where we denote $\mu_0$ to be the geodesic distance $\upmu(x,Jx)$ on $V=0$.
Introducing the variable $u = 2 U^2 + \cosh\left(\nicefrac{\pi r_+}{\ell}\right)$ we can express the above as 
\begin{eqnarray}
T(U) = \left(u -  \cosh\left(\nicefrac{\pi r_+}{\ell}\right)\right)
\frac{\left( 1 + 2 u + m\ell (u + 2) \sqrt{u^2 - 1} + m^2\ell^2 (u^2 - 1)\right)}
{\pi\ell (u^2 - 1)^{5/2} \left(u + \sqrt{u^2-1}\right)^{m\ell}} \ .
\end{eqnarray}
Clearly, the function $T(U)$ is symmetric about $U=0$. It can also be easily verified that it vanishes at $U=0$ as well as in the limit 
$U\rightarrow \pm\infty$, and is positive definite for all other values of $U$. A sketch of $T(U)$ for different values of the parameters
is depicted in Fig.\ref{tofua}.

\begin{figure}
\fbox{\includegraphics[height=3in]{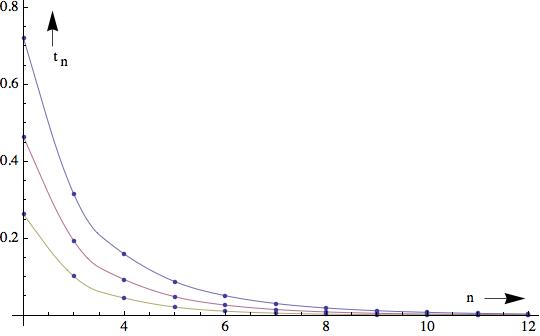}}
\caption{ Variation of $t_n$ with respect to $n$ for $r_+=1/12\pi, 1/10\pi$ and $1/8\pi$ respectively. We have set $m=0$ and $\ell=1$.
 The values of $t_n$ for a fixed $n$ decrease with increasing $r_+$.}
 \label{tofub}
\end{figure}

\begin{figure}
\fbox{\includegraphics[height=3in]{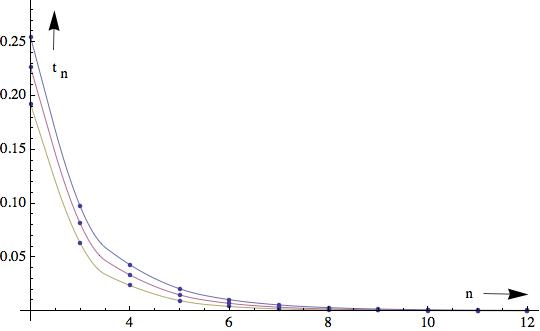}}
\caption{ Variation of $t_n$ with respect to $n$ for $m=1/2,1$ and $3/2$ respectively. We have set $r_+=1/8\pi$ and $\ell=1$.
 The values of $t_n$ for a fixed $n$ decrease with increasing $m$.}
 \label{tofuc}
\end{figure}

We now turn our attention to the average null energy for the $\mathbb{RP}^2$-geon. Here we need to sum the contribution from all the
images as described in \eqref{ads-null}, with $(\varphi-\varphi_n') = (2 n + 1) \pi$. Since $T(U)$ is positive definite, we need to choose 
anti-periodic boundary condition for the vector field $A_\rho(x)$ in the quotient space in order to violate the ANEC. The average null 
energy is now given by
\begin{equation}\label{series:one}
\langle T\rangle = - \sum_{n=0}^\infty t_n \ , 
\end{equation}
where $t_n$ is defined as (with $c_n \equiv \cosh\left(\nicefrac{(2 n + 1)\pi r_+}{\ell}\right)$):
\begin{equation}
t_n = 2 \int_{c_n}^\infty du \sqrt{\left(u -  c_n\right)}
\frac{\left( 1 + 2 u + m\ell (u + 2) \sqrt{u^2 - 1} + m^2\ell^2 (u^2 - 1)\right)}
{\pi\ell (u^2 - 1)^{5/2} \left(u + \sqrt{u^2-1}\right)^{m\ell}} \ .
\end{equation}
The integration can be exactly evaluated for certain specific values of the parameters to express it in terms of elliptic functions. The 
result is not in particular illuminating. We will instead evaluate it numerically. Without loss of generality we will set the $AdS$ radius
$\ell$ to one. The values $t_n$ for different choices of $m$ and $r_+$ are depicted in Figs. \ref{tofub} and \ref{tofuc}. 

From the numerical analysis we can see that the value of $t_n$ drops rapidly for large $n$ and the series \eqref{series:one} indeed 
converges quickly. In the following we will prove the convergence analytically. Note that we can express $t_n$ as:
\begin{equation}
t_n = \frac{2}{\pi\ell} \int_{c_n}^\infty du f(u) \frac{\sqrt{u-c_n}}{u^2 - 1} \ , 
\end{equation}
with 
\begin{equation}
f(u) =  \frac{\left( 1 + 2 u + m\ell (u + 2) \sqrt{u^2 - 1} + m^2\ell^2 (u^2 - 1)\right)}
{(u^2 - 1)^{3/2} \left(u + \sqrt{u^2-1}\right)^{m\ell}} \ .
\end{equation}
Now, observe that, for any given $r_+>0$, there exists an integer $N$ such that the function $f(u)$ as defined above takes 
value in the range $0< f(u)<1$ for $u > c_N$, for all $m\ell>0$. As a consequence of the above, if we introduce 
\begin{equation} \label{convg}
\tilde t_n = \frac{2}{\pi\ell} \int_{c_n}^\infty du \frac{\sqrt{u-c_n}}{u^2 - 1} \ , 
\end{equation}
we find $\tilde t_n>t_n$ for all $n> N$. The integration in \eqref{convg} can be evaluated exactly to find 
\begin{equation}
\tilde t_n = \frac{1}{\ell}\big(\sqrt{c_n+1} - \sqrt{c_n-1}\big) \ .
\end{equation}
Substituting $c_n = \cosh\left(\nicefrac{(2 n + 1)\pi r_+}{\ell}\right)$ in the above we find that:
\begin{equation}
\lim_{n\rightarrow\infty} \frac{\tilde t_{n+1}}{\tilde t_n} = e^{-\pi r_+/\ell} \ .
\end{equation}
Thus, the series $\sum_{n=N}^\infty\tilde t_n$ converges for any $r_+/\ell>0$. As a consequence, the series 
$\sum_{n=N}^\infty t_n$ as well as the one given in \eqref{series:one} converges giving rise to a finite value for 
$\Delta V(\infty)$.

From the numerical analysis we find that, for a fixed value of $r_+$ the opening of the wormhole is maximum when the mass of
the vector field saturates the BF bound. The size decreases as we increase $m$ and the wormhole closes in the limit $m\rightarrow\infty$.
Similarly, for fixed $m$ the opening decreases as we increase the value of $r_+$.  The size of opening remains finite for finite values 
of $m$ and $r_+$. This is in contrast to the case of scalar fields \cite{Fu:2018oaq}, where the average null energy diverges and one 
needs to adopt a regularisation process to get a finite size of the opening of the wormhole.

\section{Conclusion}

In this paper we have discussed the issue of traversability of wormholes in a quotient of the BTZ space-time by certain $\mathbb Z_2$
symmetry in the presence of massive vector fields. We have obtained the expression for the two-point function of the vector fields in 
$AdS_3$. Using this we have computed the average null energy and found that it becomes negative for appropriate choice of boundary 
conditions on the vector fields. The back reaction on the geometry then makes the wormholes traversable. It would be interesting to 
extend our analysis for quotients of $BTZ$ as well as $BTZ\times S^1$ geometries by more general discrete symmetries as well as study 
the effect of rotation on the traversability. It is also worth generalising this to study the traversability of wormholes in the presence of 
higher spin fields. Similar analysis can be carried out for higher dimensional black holes as well. More generally, recent studies have shown 
that the Euclidean wormholes play a significant role in giving a new perspective on the information loss paradox. It would be interesting to 
see whether the issue of traversability in Lorentzian wormholes such as the ones studied in the present work shed any light on this problem. 

 \bigskip\bigskip

\noindent
{\Large\bf Acknowledgement}

This work is partially supported by the DST project grant EMR/2016/001997.

\appendix

\section{The two-point function}

In this appendix we will derive the two-point function involving the field strengths corresponding 
to the gauge field $A_\rho(x)$ in $AdS_3$. Consider the Green's function
\begin{equation}
\langle A_\rho(x) A_{\sigma'}(x')\rangle =  \upalpha(\upmu) g_{\rho\sigma'}(x,x') + \upbeta(\upmu) n_\rho(x,x') n_{\sigma'}(x,x') \ .
\end{equation}
The expressions for $\upalpha(\upmu)$ and $\upbeta(\upmu)$ have been derived in \eqref{abgfmla}.
For maximally symmetric spaces, the parallel propagator $g_{\mu\nu'}(x,x')$ along the geodesic joining $x'$ to $x'$ is unique. 
It has the following properties:
\begin{eqnarray}
g_{\mu\nu'}(x,x') & = & g_{\mu\nu}(x) \ {\rm for}\  x'=x,  \label{met} \\
 g_{\mu\nu'}(x,x') & = & g_{\nu'\mu}(x',x) \label{exch} \ , \\
 g_{\mu\nu}(x) &=& g_{\mu\rho'}(x,x') g_{\nu\sigma'}(x,x') g^{\rho'\sigma'}(x') \ .  \label{asso}
 \end{eqnarray}

We need to evaluate $\langle F_{\eta\delta}(x) F_{\rho'\sigma'}(x')\rangle$. Let us first compute
$\langle \nabla_\eta A_\delta(x) \nabla_{\rho'}A_{\sigma'}(x')\rangle$. In order to do so, we need
to know the action of the covariant derivative on bitensors. This has been worked out in general
for arbitrary dimensions in \cite{Allen:1985wd}. In the following we will list the formulae relevant
for our purpose:
\begin{eqnarray}
\nabla_\rho n_\sigma(x,x') &=& A(\upmu) \big( g_{\rho\sigma}(x) - n_\rho(x,x') n_\sigma(x,x') \big) \ , \cr
\nabla_\rho n_{\sigma'}(x,x') &=& C(\upmu) \big( g_{\rho\sigma'}(x,x') + n_\rho(x,x') n_{\sigma'}(x,x') \big) \ ,  \cr
\nabla_\rho g_{\sigma\eta'}(x,x')&=& - \big(A(\upmu) + C(\upmu)\big) \big( g_{\rho\sigma}(x) n_{\eta'}(x,x')
+ g_{\rho\eta'}(x,x') \eta_\sigma(x,x') \big) \ ,
\end{eqnarray}
and analogous expressions for the derivatives with respect to $\nabla_{\rho'}$.
The functions $A(\upmu)$ and $C(\upmu)$ are defined as
\begin{equation}
A(\upmu) = \frac{1}{\ell} \coth\left(\nicefrac{\upmu}{\ell}\right) , \ {\rm and } \ 
C(\upmu) = - \frac{1}{\ell} \rm{cosech}\left(\nicefrac{\upmu}{\ell}\right) \ .
\end{equation}
In addition, note that the covariant derivative acts on an arbitrary function $f(\upmu)$ as 
$\nabla_\rho f(\upmu) = f'(\upmu)\nabla_\rho\upmu = f'(\upmu) n_\rho(x,x')$, where the prime denotes the 
derivative with respect to the argument.

Using these relations we find,
\begin{eqnarray}
\langle \nabla_\eta A_\delta(x) \nabla_{\rho'}A_{\sigma'}(x')\rangle &=&
f_1(\upmu) g_{\delta\sigma'} n_\eta n_{\rho'}
+ f_2(\upmu) g_{\delta\rho'} n_\eta n_{\sigma'} 
 +   f_3(\upmu) g_{\eta\rho'} n_\delta n_{\sigma'}  
  + f_4(\upmu) g_{\eta\sigma'} n_\delta n_{\rho'} \cr
&+& f_5(\upmu) g_{\eta\delta} n_{\rho'} n_{\sigma'} 
+  f_6(\upmu) g_{\rho'\sigma'} n_\eta n_\delta 
+ f_7(\upmu) g_{\delta\sigma'} g_{\eta\rho'} + f_8(\upmu) g_{\eta\delta} g_{\rho'\sigma'} \cr
&+& f_9(\upmu) g_{\eta\sigma'} g_{\delta\rho'}
+ f_{10}(\upmu) n_\eta n_\delta n_{\rho'} n_{\sigma'} \ ,
\end{eqnarray}
in terms of the functions $f_i(\upmu)$ which have the following expressions:
\begin{eqnarray}
&& f_1 = \alpha'' + \alpha' C , \ f_2 = C \big(\beta' - \beta (A-C)\big) - \alpha' (A+C) \cr
&& f_3 = C \big(\beta' - \beta (A-C)\big) - (A+C) \big(\beta C - \alpha (A+C)\big) \cr
&& f_4 = C  \big(\beta C - \alpha (A+C)\big) + \frac{d}{d\upmu}  \big(\beta C - \alpha (A+C)\big)  \cr
&& f_5 = \frac{d}{d\upmu}\big(\beta A - \alpha(A+C)\big) - A \big(\beta A - \alpha(A+C)\big) \cr
&& f_6 = A \big(\beta' - \beta (A-C)\big) - (A+C) \big(\alpha' + \beta C - \alpha (A+C)\big) \cr
&& f_7 = \alpha' C , f_8 = A  \big(\beta A - \alpha(A+C)\big) , f_9 = C \big(\beta C - \alpha (A+C)\big) \cr
&& f_{10} = (2 C - A) \big(\beta' - \beta (A-C)\big)  + \frac{d}{d\upmu} \big(\beta' - \beta (A-C)\big) \ . 
\end{eqnarray}
The two-point function involving the field strengths is now given by
\begin{eqnarray}
\langle F_{\eta\delta}(x) F_{\rho'\sigma'}(x')\rangle 
&=& (f_1 - f_2 + f_3 - f_4) \big(g_{\delta\sigma'} n_\eta n_{\rho'} - g_{\eta\sigma'} n_\delta n_{\rho'}
+ g_{\eta\rho'} n_\delta n_{\sigma'}  - g_{\delta\rho'} n_\eta n_{\sigma'} \big) \cr 
&+& 2 (f_9 - f_7) \big( g_{\eta\sigma'} g_{\delta\rho'} - g_{\delta\sigma'} g_{\eta\rho'} \big) \ .
\end{eqnarray}
Substituting the values of $f_i(\upmu), A(\upmu)$ and $C(\upmu)$ in the above we obtain \eqref{twofc}.

\section{The geodesic distance}

Here we will derive the expression for the geodesic distance in $AdS_3$ as well as compute the unit vectors $n_\mu(x,x')$ and $n_{\mu'}(x,x')$.
To find the geodesic distance, we consider the embedding of $AdS_3$ in $(2,2)$ Minkowski space with co-ordinates $Y^a$. The geodesic distance
$\upmu(x,x')$ is then given by
\begin{equation}
\cosh\left(\nicefrac{\upmu(x,x')}{\ell}\right) = - \frac{1}{\ell^2} \eta_{ab} Y^a(x) Y^b(x') \ .
\end{equation}
We use the Kruskal-like coordinates $x^\mu = (U,V,\varphi)$ such that
\begin{eqnarray}
Y^0 &=& \ell \frac{U+V}{1+ U V} \ , \ \ 
Y^1 = \ell \frac{1 - U V}{1 + U V} \cosh\left(\nicefrac{r_+\varphi}{\ell}\right) \ ,  \cr
Y^2 &=& \ell \frac{U-V}{1+ U V}  \ ,  \ \
Y^3 = \ell \frac{1 - U V}{1 + U V} \sinh\left(\nicefrac{r_+\varphi}{\ell}\right) \ .
\end{eqnarray}
In terms of these coordinates, the geodesic distance $\upmu(x,x')$ between points $x$ and $x'$  is 
\begin{equation}
\cosh\left(\nicefrac{\upmu}{\ell}\right)  = \frac{
\Big( 2 \left( U V' + V U'\right) + \left(1 - U V\right) \left(1 - U' V'\right) \cosh\left(\nicefrac{r_+(\varphi - \varphi')}{\ell}\right)\Big)
}{\left(1 + U V\right) \left(1 + U' V'\right)} \ . 
\end{equation}

It is now straightforward to compute the components of the unit tangents $n_\rho(x,x')$ and $n_{\rho'}(x,x')$.
We find
\begin{eqnarray}
n_U(x,x') &=& \frac{2\ell}{\sinh\left(\nicefrac{\upmu}{\ell}\right)}
\frac{\Big(V' - V^2 U' - \left(1 - U' V'\right) V \cosh\left(\nicefrac{r_+(\varphi-\varphi')}{\ell}\right)\Big)}
{\left(1 + U V\right)^2 \left(1 + U' V'\right)} \cr
n_V(x,x') &=& \frac{2\ell}{\sinh\left(\nicefrac{\upmu}{\ell}\right)}
\frac{\Big( U' - U^2 V' - (1 - U' V') U \cosh\left(\nicefrac{r_+(\varphi-\varphi')}{\ell}\right)
\Big)}{\left(1 + U V\right)^2 \left(1 + U' V'\right)} \cr
n_\varphi(x,x') &=& \frac{r_+}{\sinh\left(\nicefrac{\upmu}{\ell}\right)}
\frac{\Big(  \left(1 - U V \right)  \left(1 - U' V'\right)\sinh\left(\nicefrac{r_+(\varphi - \varphi')}{\ell}\right)\Big)}
{\left(1 + U V\right) \left(1 + U' V'\right)} \ . 
\end{eqnarray}
Likewise, components of the unit tangent  $n_{\rho'}(x,x') = \nabla_{\rho'}\upmu(x,x')$ at $x'$ are
\begin{eqnarray}
n_{U'}(x,x') &=& \frac{2\ell}{\sinh\left(\nicefrac{\upmu}{\ell}\right)}
\frac{\Big(V - U V'^2  - \left(1 - U V\right) V' \cosh\left(\nicefrac{r_+(\varphi-\varphi')}{\ell}\right)\Big)}
{\left(1 + U V\right) \left(1 + U' V'\right)^2} \cr
n_{V'}(x,x') &=& \frac{2\ell}{\sinh\left(\nicefrac{\upmu}{\ell}\right)}
\frac{\Big( U - V U'^2  - (1 - U V ) U' \cosh\left(\nicefrac{r_+(\varphi-\varphi')}{\ell}\right)
\Big)}{\left(1 + U V\right) \left(1 + U' V'\right)^2} \cr
n_{\varphi'}(x,x') &=& -  \frac{r_+}{\sinh\left(\nicefrac{\upmu}{\ell}\right)}
\frac{\Big(  \left(1 - U V \right)  \left(1 - U' V'\right)\sinh\left(\nicefrac{r_+(\varphi - \varphi')}{\ell}\right)\Big)}
{\left(1 + U V\right) \left(1 + U' V'\right)} \ . 
\end{eqnarray}
Setting $x'=Jx$, we find 
\begin{eqnarray}
n_U(x,Jx) &=& \frac{2\ell \left(U - V^3 - V (1 - U V) \cosh\left(\nicefrac{k \pi r_+}{\ell}\right)\right)}
{\sinh\left(\nicefrac{\upmu}{\ell}\right) \left(1 + U V\right)^3} \cr
n_V(x,Jx) &=& \frac{2\ell \left(V - U^3 - U (1 - U V) \cosh\left(\nicefrac{k \pi r_+}{\ell}\right)\right)}
{\sinh\left(\nicefrac{\upmu}{\ell}\right) \left(1 + U V\right)^3} \cr
n_\varphi(x,Jx) &=& -  \frac{r_+ \left(1 - U V\right)^2 \sinh\left(\nicefrac{k \pi r_+}{\ell}\right)}
{\sinh\left(\nicefrac{\upmu}{\ell}\right) \left(1 + U V\right)^2} \ , 
\end{eqnarray}
and 
\begin{eqnarray}
n_{U'}(x,Jx) &=& \frac{2\ell \left(V - U^3 - U (1 - U V) \cosh\left(\nicefrac{k \pi r_+}{\ell}\right)\right)}
{\sinh\left(\nicefrac{\upmu}{\ell}\right) \left(1 + U V\right)^3} \cr
n_{V'}(x,Jx) &=& \frac{2\ell \left(U - V^3 - V (1 - U V) \cosh\left(\nicefrac{k \pi r_+}{\ell}\right)\right)}
{\sinh\left(\nicefrac{\upmu}{\ell}\right) \left(1 + U V\right)^3} \cr
n_{\varphi'}(x,Jx) &=&  \frac{r_+ \left(1 - U V\right)^2 \sinh\left(\nicefrac{k \pi r_+}{\ell}\right)}
{\sinh\left(\nicefrac{\upmu}{\ell}\right) \left(1 + U V\right)^2} \ .
\end{eqnarray}

\section{The parallel propagator}

In this appendix we will analyse the behaviour of the parallel propagator on the $V=0$ surface. The parallel propagator $g_{\mu\nu'}(x,x')$ 
is a linear map which parallel transports vectors $V^{\rho'}(x')$ at the point $x'$ to 
$V^\rho(x)$ at the point $x$:
\begin{equation}
g_{\mu\nu'}(x,x') V^{\nu'}(x') = V_\mu(x) \ .
\end{equation}
The equation of motion for the parallel propagator can be obtained from the equation for parallel transport of a vector and is given by:
\begin{equation}\label{ppgvz}
\frac{d}{d\lambda} {g^\mu}_{\nu'}(x,x') + \Gamma^\mu_{\nu\rho} \frac{dx^\nu}{d\lambda}{g^\rho}_{\nu'}(x,x') = 0 \ .
\end{equation}
It depends on the path $x^\mu(\lambda)$ and has the following formal solution in terms of the path-ordered exponential  \cite{Carroll:1997ar}:
\begin{equation}
{g^\mu}_{\nu'}(x,x') = {\cal P} \exp\left(-\int_{\lambda'}^\lambda  \Gamma^\mu_{\sigma\nu} 
\frac{dx^\sigma}{d\lambda} d\lambda \right) \ .
\end{equation}

In the following we will analyse \eqref{ppgvz} for $AdS_3$ in detail. The non-vanishing components of the affine connection are listed as:
\begin{eqnarray} 
& & \Gamma^U_{UU} = - \frac{2 V}{1 + U V} \ ,  \  \Gamma^U_{\varphi\varphi} = - \frac{r_+^2 U (1 - U V)}{\ell^2 (1 + U V)} \ , \cr
& & \Gamma^V_{VV} = - \frac{2 U}{1 + U V} \ ,  \
  \Gamma^V_{\varphi\varphi} = - \frac{r_+^2 V (1 - U V)}{\ell^2 (1 + U V)} \ , \cr
& &\Gamma^\varphi_{\varphi U} = - \frac{2 V}{1 - U^2 V^2} \ , \  \Gamma^\varphi_{\varphi V} = - \frac{2 U}{1 - U^2 V^2} \ .
\end{eqnarray}
The equation of motion \eqref{ppgvz} for the parallel propagator now becomes
\begin{eqnarray} \label{c5}
&& \frac{d}{d\lambda} {g^U}_{U'}- \frac{2 V}{1 + U V} \frac{dU}{d\lambda} {g^U}_{U'} 
- \frac{r_+^2 U (1 - U V)}{\ell^2 (1 + U V)} \frac{d\varphi}{d\lambda} {g^\varphi}_{U'} = 0 , \cr
&& \frac{d}{d\lambda} {g^U}_{V'} - \frac{2 V}{1 + U V} \frac{dU}{d\lambda} {g^U}_{V'} 
- \frac{r_+^2 U (1 - U V)}{\ell^2 (1 + U V)} \frac{d\varphi}{d\lambda} {g^\varphi}_{V'} = 0 , \cr
&& \frac{d}{d\lambda} {g^U}_{\varphi'} - \frac{2 V}{1 + U V} \frac{dU}{d\lambda} {g^U}_{\varphi'} 
- \frac{r_+^2 U (1 - U V)}{\ell^2 (1 + U V)} \frac{d\varphi}{d\lambda} {g^\varphi}_{\varphi'} = 0 , \cr
&& \frac{d}{d\lambda} {g^V}_{U'}- \frac{2 U}{1 + U V} \frac{dV}{d\lambda} {g^V}_{U'} 
- \frac{r_+^2 V (1 - U V)}{\ell^2 (1 + U V)} \frac{d\varphi}{d\lambda} {g^\varphi}_{U'} = 0 , \cr
&& \frac{d}{d\lambda} {g^V}_{V'}- \frac{2 U}{1 + U V} \frac{dV}{d\lambda} {g^V}_{V'} 
- \frac{r_+^2 V (1 - U V)}{\ell^2 (1 + U V)} \frac{d\varphi}{d\lambda} {g^\varphi}_{V'} = 0 , \cr
&& \frac{d}{d\lambda} {g^V}_{\varphi'}- \frac{2 U}{1 + U V} \frac{dV}{d\lambda} {g^V}_{\varphi'} 
- \frac{r_+^2 V (1 - U V)}{\ell^2 (1 + U V)} \frac{d\varphi}{d\lambda} {g^\varphi}_{\varphi'} = 0 , \cr
&& \frac{d}{d\lambda} {g^\varphi}_{U'}- \frac{2}{1 - U^2 V^2} \left( 
V \frac{d\varphi}{d\lambda} {g^U}_{U'}
+  U \frac{d\varphi}{d\lambda} {g^V}_{U'}
+ \frac{d(U V)}{d\lambda} {g^\varphi}_{U'}\right) = 0 \cr
&& \frac{d}{d\lambda} {g^\varphi}_{V'}- \frac{2}{1 - U^2 V^2} \left( 
V \frac{d\varphi}{d\lambda} {g^U}_{V'}
+  U \frac{d\varphi}{d\lambda} {g^V}_{V'}
+ \frac{d(U V)}{d\lambda} {g^\varphi}_{V'}\right) = 0 \cr
&& \frac{d}{d\lambda} {g^\varphi}_{\varphi'}- \frac{2}{1 - U^2 V^2} \left( 
V \frac{d\varphi}{d\lambda} {g^U}_{\varphi'}
+  U \frac{d\varphi}{d\lambda} {g^V}_{\varphi'}
+ \frac{d(U V)}{d\lambda} {g^\varphi}_{\varphi'}\right) = 0 
\end{eqnarray}

The above equations need to be analysed along the geodesic:
\begin{equation}
\frac{d^2x^\rho}{d\lambda^2} + \Gamma^\rho_{\eta\sigma}
\frac{dx^\eta}{d\lambda}\frac{dx^\sigma}{d\lambda}=0 .
\end{equation}
For the present case this becomes
\begin{eqnarray}
 \frac{d^2U}{d\lambda^2} - \frac{2 V}{1 + U V} \left(\frac{dU}{d\lambda}\right)^2 - \frac{r_+^2 U (1 - U V)}{\ell^2 (1 + U V)} \left(\frac{d\varphi}{d\lambda}\right)^2 &=& 0 \cr
 \frac{d^2V}{d\lambda^2} - \frac{2 U}{1 + U V} \left(\frac{dV}{d\lambda}\right)^2 - \frac{r_+^2 V (1 - U V)}{\ell^2 (1 + U V)} \left(\frac{d\varphi}{d\lambda}\right)^2 &=& 0 \cr
 \frac{d^2\varphi}{d\lambda^2} - \frac{4}{1 - U^2 V^2} \frac{d(UV)}{d\lambda} \frac{d\varphi}{d\lambda} &=& 0
\end{eqnarray}

We will now choose $U$ to be the affine parameter. Setting $V=0$ in the geodesic equation, we can easily see that $\frac{d\varphi}{dU}=0$. 
Using this in \eqref{c5}, we find that the term $\Gamma^\mu_{\nu\rho} \frac{dx^\nu}{d\lambda}{g^\rho}_{\nu'}(x,x')$ vanishes identically and 
hence ${g^\rho}_{\sigma'}(x,x')$ is independent of $U$ on the $V=0$ surface.


\begin{thebibliography}{99}


%\cite{Morris:1988cz}
\bibitem{Morris:1988cz}
M.~S.~Morris and K.~S.~Thorne,
%``Wormholes in space-time and their use for interstellar travel: A tool for teaching general relativity,''
Am. J. Phys. \textbf{56}, 395-412 (1988)
doi:10.1119/1.15620
%1378 citations counted in INSPIRE as of 04 Aug 2020  


%\cite{Hochberg:1998ha}
\bibitem{Hochberg:1998ha}
D.~Hochberg and M.~Visser,
%``Dynamic wormholes, anti-trapped surfaces, and energy conditions,''
Phys. Rev. D \textbf{58}, 044021 (1998)
doi:10.1103/PhysRevD.58.044021
[arXiv:gr-qc/9802046 [gr-qc]].
%186 citations counted in INSPIRE as of 04 Aug 2020

%\cite{Morris:1988tu}
\bibitem{Morris:1988tu}
M.~S.~Morris, K.~S.~Thorne and U.~Yurtsever,
%``Wormholes, Time Machines, and the Weak Energy Condition,''
Phys. Rev. Lett. \textbf{61}, 1446-1449 (1988)
doi:10.1103/PhysRevLett.61.1446
%949 citations counted in INSPIRE as of 04 Aug 2020

%\cite{Visser:2003yf}
\bibitem{Visser:2003yf}
M.~Visser, S.~Kar and N.~Dadhich,
%``Traversable wormholes with arbitrarily small energy condition violations,''
Phys. Rev. Lett. \textbf{90}, 201102 (2003)
doi:10.1103/PhysRevLett.90.201102
[arXiv:gr-qc/0301003 [gr-qc]].
%242 citations counted in INSPIRE as of 04 Aug 2020

%\cite{Graham:2007va}
\bibitem{Graham:2007va}
N.~Graham and K.~D.~Olum,
%``Achronal averaged null energy condition,''
Phys. Rev. D \textbf{76}, 064001 (2007)
doi:10.1103/PhysRevD.76.064001
[arXiv:0705.3193 [gr-qc]].
%54 citations counted in INSPIRE as of 04 Aug 2020

%\cite{Wall:2009wi}
\bibitem{Wall:2009wi}
A.~C.~Wall,
%``Proving the Achronal Averaged Null Energy Condition from the Generalized Second Law,''
Phys. Rev. D \textbf{81}, 024038 (2010)
doi:10.1103/PhysRevD.81.024038
[arXiv:0910.5751 [gr-qc]].
%31 citations counted in INSPIRE as of 04 Aug 2020

%\cite{Kelly:2014mra}
\bibitem{Kelly:2014mra}
W.~R.~Kelly and A.~C.~Wall,
%``Holographic proof of the averaged null energy condition,''
Phys. Rev. D \textbf{90}, no.10, 106003 (2014)
doi:10.1103/PhysRevD.90.106003
[arXiv:1408.3566 [gr-qc]].
%37 citations counted in INSPIRE as of 04 Aug 2020

%\cite{Friedman:1993ty}
\bibitem{Friedman:1993ty}
J.~L.~Friedman, K.~Schleich and D.~M.~Witt,
%``Topological censorship,''
Phys. Rev. Lett. \textbf{71}, 1486-1489 (1993)
doi:10.1103/PhysRevLett.71.1486
[arXiv:gr-qc/9305017 [gr-qc]].
%367 citations counted in INSPIRE as of 05 Aug 2020

%%%%\cite{Gao:2000ga}
%%%\bibitem{Gao:2000ga}
%%%S.~Gao and R.~M.~Wald,
%%%%``Theorems on gravitational time delay and related issues,''
%%%Class. Quant. Grav. \textbf{17}, 4999-5008 (2000)
%%%doi:10.1088/0264-9381/17/24/305
%%%[arXiv:gr-qc/0007021 [gr-qc]].
%%%%137 citations counted in INSPIRE as of 05 Aug 2020

%\cite{Galloway:1999bp}
\bibitem{Galloway:1999bp}
G.~J.~Galloway, K.~Schleich, D.~M.~Witt and E.~Woolgar,
%``Topological censorship and higher genus black holes,''
Phys. Rev. D \textbf{60}, 104039 (1999)
doi:10.1103/PhysRevD.60.104039
[arXiv:gr-qc/9902061 [gr-qc]].
%151 citations counted in INSPIRE as of 05 Aug 2020

%\cite{Gao:2016bin}
\bibitem{Gao:2016bin}
P.~Gao, D.~L.~Jafferis and A.~C.~Wall,
%``Traversable Wormholes via a Double Trace Deformation,''
JHEP \textbf{12}, 151 (2017)
doi:10.1007/JHEP12(2017)151
[arXiv:1608.05687 [hep-th]].
%170 citations counted in INSPIRE as of 05 Aug 2020

%\cite{Caceres:2018ehr}
\bibitem{Caceres:2018ehr}
E.~Caceres, A.~S.~Misobuchi and M.~L.~Xiao,
%``Rotating traversable wormholes in AdS,''
JHEP \textbf{12}, 005 (2018)
doi:10.1007/JHEP12(2018)005
[arXiv:1807.07239 [hep-th]].
%18 citations counted in INSPIRE as of 05 Aug 2020

%\cite{Maldacena:2018lmt}
\bibitem{Maldacena:2018lmt}
J.~Maldacena and X.~L.~Qi,
%``Eternal traversable wormhole,''
[arXiv:1804.00491 [hep-th]].
%136 citations counted in INSPIRE as of 05 Aug 2020

%\cite{Maldacena:2018gjk}
\bibitem{Maldacena:2018gjk}
J.~Maldacena, A.~Milekhin and F.~Popov,
%``Traversable wormholes in four dimensions,''
[arXiv:1807.04726 [hep-th]].
%53 citations counted in INSPIRE as of 05 Aug 2020

%\cite{Fu:2018oaq}
\bibitem{Fu:2018oaq}
Z.~Fu, B.~Grado-White and D.~Marolf,
%``A perturbative perspective on self-supporting wormholes,''
Class. Quant. Grav. \textbf{36}, no.4, 045006 (2019)
doi:10.1088/1361-6382/aafcea
[arXiv:1807.07917 [hep-th]].
%23 citations counted in INSPIRE as of 05 Aug 2020

%\cite{Marolf:2019ojx}
\bibitem{Marolf:2019ojx}
D.~Marolf and S.~McBride,
%``Simple Perturbatively Traversable Wormholes from Bulk Fermions,''
JHEP \textbf{11}, 037 (2019)
doi:10.1007/JHEP11(2019)037
[arXiv:1908.03998 [hep-th]].
%3 citations counted in INSPIRE as of 05 Aug 2020

%\cite{Maldacena:2001kr}
\bibitem{Maldacena:2001kr}
J.~M.~Maldacena,
%``Eternal black holes in anti-de Sitter,''
JHEP \textbf{04}, 021 (2003)
doi:10.1088/1126-6708/2003/04/021
[arXiv:hep-th/0106112 [hep-th]].
%943 citations counted in INSPIRE as of 24 Aug 2020

%\cite{Louko:1998hc}
\bibitem{Louko:1998hc}
J.~Louko and D.~Marolf,
%``Single exterior black holes and the AdS / CFT conjecture,''
Phys. Rev. D \textbf{59}, 066002 (1999)
doi:10.1103/PhysRevD.59.066002
[arXiv:hep-th/9808081 [hep-th]].
%60 citations counted in INSPIRE as of 25 Aug 2020


%\cite{Allen:1985wd}
\bibitem{Allen:1985wd}
B.~Allen and T.~Jacobson,
%``Vector Two Point Functions in Maximally Symmetric Spaces,''
Commun. Math. Phys. \textbf{103}, 669 (1986)
doi:10.1007/BF01211169
%274 citations counted in INSPIRE as of 08 Aug 2020


%\cite{Carroll:1997ar}
\bibitem{Carroll:1997ar}
S.~M.~Carroll,
%``Lecture notes on general relativity,''
[arXiv:gr-qc/9712019 [gr-qc]].
%154 citations counted in INSPIRE as of 25 Aug 2020











\end{thebibliography}
\end{document}